\documentclass[11pt]{article}

\usepackage{geometry}  
\usepackage[T1]{fontenc}  
\usepackage[utf8x]{inputenc}  
\usepackage{lineno}  
\usepackage{fancyhdr}  
\usepackage{enumitem}  
\usepackage{hyperref}  
\usepackage{titling}  
\usepackage[sort&compress]{natbib}  
\usepackage{mathtools}  
\usepackage{titlesec}  
\usepackage{lastpage}  

\usepackage[english]{babel}  
\usepackage{amsmath}  
\usepackage{amsfonts}  
\usepackage{amssymb}  
\usepackage{wasysym}  
\usepackage{bbm}  
\usepackage{array}  
\usepackage{xr}  
\usepackage{verbatim} 
\usepackage{float} 

\hypersetup{%
  pdfborderstyle={/S/U/W 1}
}

\setcitestyle{square,numbers}
\titleformat{name=\section}{\normalfont\Large\bfseries}{}{0pt}{}
\titleformat{name=\subsection}{\normalfont\large\bfseries}{}{0pt}{}
\titleformat{name=\subsubsection}{\normalfont\normalsize\bfseries}{}{0pt}{}

\newcommand{\email}[1]{\href{mailto:{#1}}{{#1}}}

\newcommand{\keywords}[1]{\textbf{Keywords}: {#1}}

\newcommand{\optincludegraphics}[2][{}]{\includegraphics[{#1}]{{#2}}}

\newcommand{\optinput}[1]{}

\newcommand{\capt}[2][]{\caption[{#1}]{\textbf{{#1}}\newline{#2}}}

\newtagform{brackets}{[}{]}
\usetagform{brackets}

\newcommand{\thejournal}[1]{NMR in Biomedicine}

\title{Application of a k-Space Interpolating Artificial Neural Network to In-Plane Accelerated Simultaneous Multislice Imaging}

\begin{document}

\begin{titlepage}
{\noindent\LARGE\bf \thetitle}

\bigskip

\begin{flushleft}\large
	Nikolai J. Mickevicius\textsuperscript{1,{*}},
	Eric S. Paulson\textsuperscript{1,2},
	L. Tugan Muftuler\textsuperscript{3,4},
	Andrew S. Nencka\textsuperscript{2,4}
\end{flushleft}

\bigskip

\noindent
\begin{enumerate}[label=\textbf{\arabic*}]
\item Department of Radiation Oncology, Medical College of Wisconsin, Milwaukee, WI, USA
\item Department of Radiology, Medical College of Wisconsin, Milwaukee, WI, USA
\item Department of Neurosurgery, Medical College of Wisconsin, Milwaukee, WI, USA
\item Center for Imaging Research, Medical College of Wisconsin, Milwaukee, WI, USA
\end{enumerate}

\bigskip


\textbf{*} Corresponding author:

\indent\indent
\begin{tabular}{>{\bfseries}rl}
Name		& Nikolai J. Mickevicius													\\
Department	& Department of Radiation Oncology													\\
Institute	& Medical College of Wisconsin														\\
Address 	& 8701 Watertown Plank Road														\\
			& Milwaukee, WI 53226														\\
            & USA														\\
E-mail		& \email{nmickevicius@mcw.edu}											\\
\end{tabular}

\vfill





\end{titlepage}

\pagebreak

\begin{abstract}

\textbf{Purpose}: The goal of this work is to extend the capabilities of RAKI, a k-space interpolating neural network, to reconstruct high-quality images from in-plane accelerated simultaneous multislice imaging acquisitions. This method is referred to as slice-RAKI.

\textbf{Methods}: A three-layer convolutional neural network was designed to output k-space signals for separate slices given the input of a multicoil slice-aliased k-space. The output of the slice-interpolation network is passed into a separate in-plane interpolating network for each slice. The proposed framework was tested in retrospective acceleration experiments in vivo, and in prospectively accelerated phantom and in vivo experiments.  

\textbf{Results}: The neural network interpolation based reconstruction quantitatively outperforms conventional parallel imaging reconstruction algorithms for all tested in-plane and simultaneous multislice acceleration factors. Visually, the neural network reconstructions are of superior quality compared to parallel imaging reconstructions in prospectively accelerated acquisitions.

\textbf{Conclusion}: Slice-RAKI provides a patient specific neural network based non-linear reconstruction which improves image quality compared with conventional linear parallel imaging algorithms. It could find use in MR-guided interventions such as MR-guided radiation therapy for use in rapid real-time cine imaging for motion monitoring. 

\end{abstract}

\bigskip
\keywords{simultaneous multislice, SMS, interpolation, deep learning, neural network, CNN}

\pagebreak

\section{Introduction}

Precisely delivering radiation to abdominal or thoracic tumors is challenged by the substantial motions that can occur within each treatment fraction. Respiration, peristalsis, cardiac pulsations, organ filling, drifting, and coughing can all contribute to uncertainties \citep{physioText}. Larger margins encompassing both targets and nearby organs at risk (OAR) are often employed to account for these uncertainties.  Magnetic resonance-guided radiation therapy (MR-gRT) has the potential to increase the precision of radiation delivery by imaging and segmenting the anatomy in real-time, facilitating gating of the radiation beam or adjustment of beam apertures to track the target. Either of these motion management methods would enable margins to be reduced, sparing more healthy tissue while precisely irradiating the target.

High frame rate cine imaging can be performed in real-time to track the target and gate the beam based on its position. Current implementations of real-time cine imaging on commercially available MR-gRT systems include imaging a single sagittal slice or interleaved sagittal, coronal, and axial imaging\citep{interleavedCine}. Choice of one technique results in the classic spatiotemporal resolution tradeoffs.  In the single sagittal slice case, a high frame rate is able to detect abrupt, irregular motions like coughing. However, with only one slice, it is difficult to track through-plane motion of the target and OAR. Use of interleaved orthogonal cine imaging planes improves motion estimation in three dimensions. However, the frame rate will be reduced compared to the single slice acquisition. In either case, the ability to speed up the acquisition of a single slice, or to acquire multiple slices at the same frame rate, is highly desired. Several studies aiming to increase the spatiotemporal resolution of cine imaging for MR-gRT have been performed\citep{sopi,ktsopi}.

Magnetic resonance imaging is an inherently slow imaging modality. The speed at which a 2D MR image can be acquired involves balancing hardware constraints (e.g. maximum gradient amplitude and slew rate), physiological constraints (e.g. peripheral nerve stimulation and specific absorption rate), image contrast, and the signal-to-noise ratio (SNR) of the reconstructed images\cite{bernsteinBook}. A plethora of methods have been developed to accelerate MR acquisition. Namely, parallel imaging methods\cite{parallelImagingReview} have been used ubiquitously in the past two decades as they have been made commercially available by scanner vendors.

Parallel imaging (PI) methods\citep{smash,sense,grappa} employ linear reconstruction techniques to remove aliasing artifacts caused by sampling k-space below the Nyquist limit. PI exploits redundancies in the image data measured with multiple radio-frequency (RF) coils, each with a unique spatially varying sensitivity profile. These methods can be performed within the image domain\citep{sense,cgsense}, or as interpolation operations within the k-space domain\citep{smash,grappa}. While the former approaches are optimal assuming explicit and accurate coil sensitivities\citep{espirit}, k-space-based methods do not rely on precise computation of the coil sensitivity profiles for each of the RF coils. 

Regardless of the domain in which a parallel imaging algorithm operates, the least squares solution is dependent on obtaining accurate calibration data and favorable coil geometries. In most cases, when no further regularization is employed, PI algorithms will fail due to noise enhancement and/or residual aliasing artifacts as the acceleration factor approaches the number of receive RF coils\citep{sense}. For clinical 2D imaging, acceleration factors greater than 4 are rarely seen. Compressed sensing has showed enormous potential to improve the quality of images reconstructed from highly accelerated acquisitions by enforcing sparsity in a certain transform domain (e.g. wavelet or finite difference domains)\citep{cs}. However, solutions to compressed sensing reconstructions can only be obtained iteratively and cannot be computed fast enough for real-time imaging applications. Even with highly parallelized GPU-optimized reconstructions, latencies on the order of ~0.3 seconds can be observed\citep{realTimeCS}, which may be too long for precise target location prediction during abdominal radiotherapy\citep{motionPrediction}. In order to robustly monitor and track the position of tumors during MR-gRT, a non-iterative reconstruction algorithm is desired to reconstruct highly accelerated imaging. 

The use of simultaneous multislice (SMS) imaging has been proposed to improve cine imaging during MR-gRT\citep{sopi,radialCaipiCine,particleFilter}. By monitoring the anatomy in multiple parallel planes, more precise through-plane motion estimates can be made compared to single-slice cine imaging. Parallel imaging algorithms have also been adapted to separate aliased slices in SMS acquisitions\citep{larkman}. As with in-plane acceleration, the PI problem is better posed when the differences in coil sensitivities of overlapped slices are maximized. Methods to produce a controlled aliasing of the simultaneously excited slices, dubbed CAIPIRINHA, introduce inter-slice FOV shifts which will increase the variation in coil sensitivities in overlapped voxels\citep{caipi}. These shifts can be generated via RF phase cycling or gradient blips along the slice-select direction\citep{caipi,blippedCaipi}.

Deep learning (DL) has recently gained traction in the MR image reconstruction field\citep{variationalNetwork,domainAdaptation,residualLearningMR,kikiNet}. Like parallel imaging, DL networks can learn to remove aliasing artifacts in the image domain or interpolate undersampled data in the k-space domain. DL-based reconstructions typically involve an extensive training dataset from a multitude of patients and anatomical sites. Recently, the Robust Artificial-neural-networks for K-space Interpolation (RAKI) method was introduced to calculate a scan-specific neural network to interpolate skipped phase encoding samples\citep{raki}. Analogous to calculating linear interpolation weights from a fully-sampled portion of k-space in the GRAPPA parallel imaging method, RAKI calculates many kernels in a non-linear 3-layer convolutional neural network. RAKI is different from most of the other DL reconstruction methods, and similar to KIKI-net\citep{kikiNet}, in that it utilizes the measured k-space data from all RF coils, rather than a coil-combined k-space. RAKI was shown to improve the reconstruction results when compared with GRAPPA at high acceleration factors. 

In the present study, the originally proposed RAKI method is extended for SMS imaging. Its performance relative to conventional parallel imaging algorithms is tested quantitatively at a variety of acceleration factors retrospectively. Finally, the RAKI method is used to reconstruct prospectively accelerated phantom data and in vivo SMS cine imaging data.

\section{Methods}

\subsection{Slice-RAKI}
In the slice-GRAPPA algorithm, data from simultaneously excited slices are separated from the slice-collapsed k-space\citep{blippedCaipi}. Similar to standard GRAPPA, linear convolution kernels are used to perform the slice separation. Unlike GRAPPA, however, slice-GRAPPA interpolates entirely new sets of k-space for each slice (i.e. no directly measured data remains in the k-space matrices for any SMS slice). The linear convolution kernels in slice-GRAPPA are calculated using linear least squares methods given fully sampled data for each slice (See Setsompop et. al. 2012)\citep{blippedCaipi}. As shown in Akçakaya et. al. (2018)\citep{raki}, the use of non-linear k-space interpolation (RAKI) improves the resiliency to noise in both the training and accelerated data relative to GRAPPA. The RAKI method was extended, as subsequently described, to perform slice unaliasing via non-linear k-space interpolation analogous to slice-GRAPPA. 

\subsection{Network Description}
\begin{figure}[H] 
\optincludegraphics[width=\textwidth]{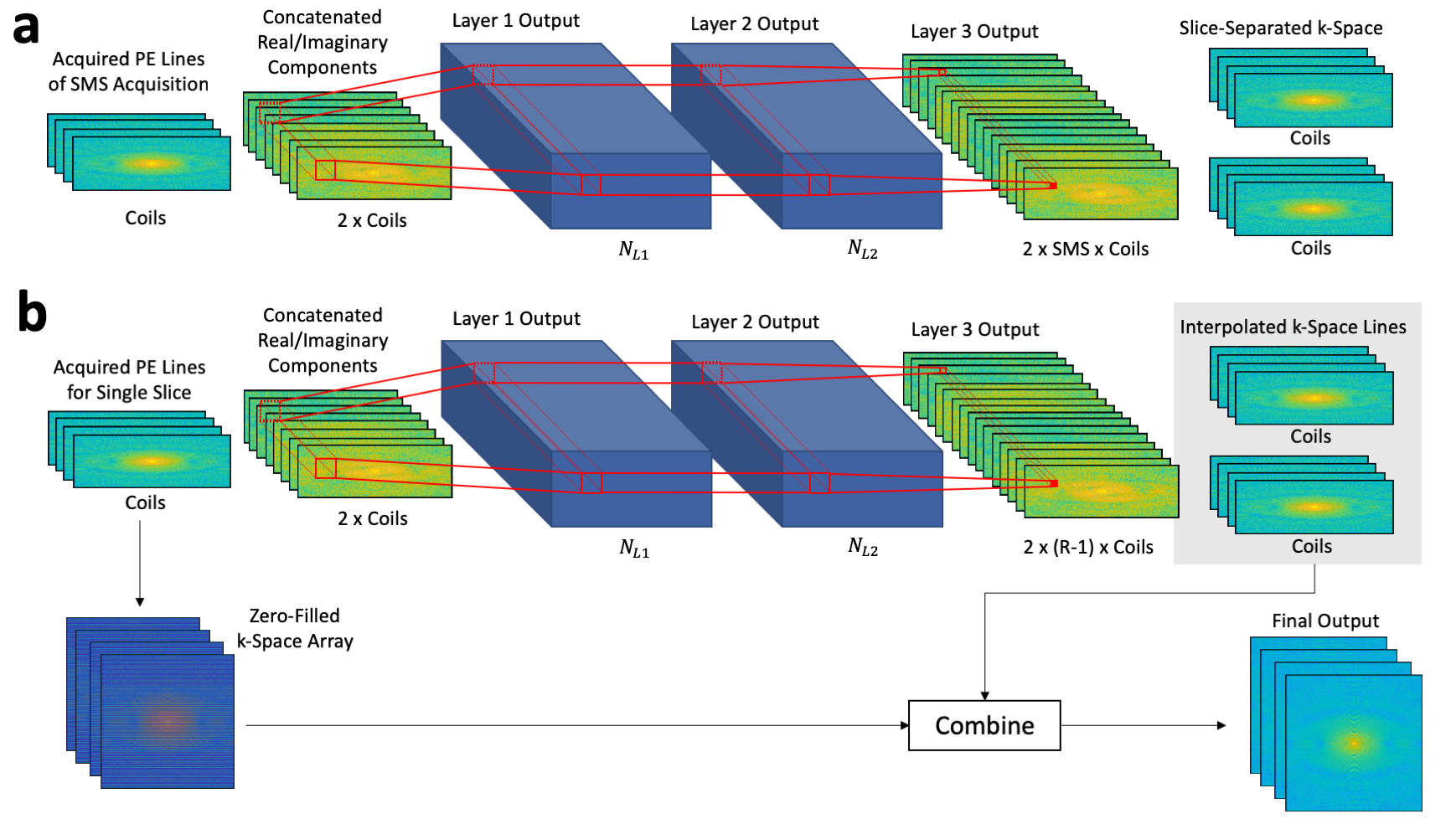}
\capt[Slice-RAKI Network.]{Reconstruction of in-plane accelerated simultaneous multislice MR data using RAKI. (a) Slice-RAKI uses a three-layer CNN to separate the slice-aliased k-space data into individual SMS slices. (b) For each slice output from the slice-RAKI network, the standard RAKI network employs a three-layer CNN to interpolate (R-1) skipped phase encode lines. The input to RAKI and the outputs are combined to create a final interpolated k-space.}
\label{fig:network}
\end{figure}

RAKI performs multi-coil k-space interpolation using a simple heuristically-determined convolution neural network (CNN) with three layers. A depiction of RAKI for slice unaliasing and in-plane interpolation can be found in Figure \ref{fig:network}. Due to implementation barriers, the real and imaginary components of the k-space data are included as extra “channels” in the input since TensorFlow does not easily accommodate complex-valued data types. For slice-RAKI, the data input to the network is the slice-aliased data uniformly sampled along the phase-encoding direction. The output of the slice-RAKI network is the slice-separated k-space signals at each measured phase encoding line. If additional in-plane acceleration was employed, the output of the slice-RAKI network is fed into a standard RAKI network. The outputs of the standard RAKI network are the interpolated phase encode lines which were skipped during the acquisition. 

\subsection{Implementation Details}
The RAKI networks were implemented in TensorFlow and were trained using random patches of a single fully-sampled k-space dataset from each of the separate SMS slices. The kernel dimensions (height x width) are 5x5, 1x1, and 3x3 for each layer of the slice-RAKI network, respectively. For the standard RAKI network, the kernel dimensions were 4x5, 1x1, and 2x3 for each layer of the standard RAKI network, respectively. In both networks, the rectified linear unit (ReLU) activation was applied only after the second layer. The number of channels at the outputs of the first and second layer were $N_{L1} = 128$ and $N_{L2} = 128$ for both networks. The number of training epochs for the slice- and standard-RAKI networks were only 500 and 100, respectively, given the relative simplicity of the network architecture. A mini-batch size of 8 regions of interest of size 24 pixels x 24 pixels x (2 x the number of coils) was used within each epoch. The gradient descent optimizer with a learning rate of 0.001 was used to learn kernel weights while minimizing mean squared error (MSE). Normalization was applied as recommended in Akçakaya et. al. (2018)\citep{raki}. No biases were learned. A description of the operations performed by each of the three layers of the RAKI network are shown below. The first layer (Eq. 1) convolves the undersampled measured k-space with filters of size $N_{1,PE}$ x $N_{1,RO}$ x ($2N_C$) x $N_{L1}$, where $N_{1,PE}$ and $N_{1,RO}$ are the number of phase encode lines and readout points input to the network, respectively. $N_C$ represents the number of coils. The second layer (Eq. 2) convolves the output of layer one with convolution filters of size $N_{1,PE}$ x $N_{1,RO}$ x $N_{L1}$ x $N_{L2}$ followed by a ReLU operation. Finally, layer three (Eq. 3) convolves the output of layer two with filters of size $N_{1,PE}$ x $N_{1,RO}$ x $N_{L2}$ x $N_{OUT}$. The output layer sizes are given by $N_{OUT}=2\cdot\left(R-1\right)\cdot N_C$ and $N_{OUT}=2\cdot\left(SMS\right)\cdot N_C$ for the in-plane and slice RAKI networks, respectively.
\begin{equation}
    F_1(\mathbf{s})=\mathbf{w}_1*\mathbf{s}
\end{equation}
\begin{equation}
    F_2(\mathbf{s})=ReLU\left(\mathbf{w}_2*\mathbf{s}\right)
\end{equation}
\begin{equation}
    F_3(\mathbf{s})=\mathbf{w}_3*\mathbf{s}
\end{equation}

Despite the fact that no biases were learned in the CNN, it was observed that the slice-RAKI network can introduce small DC offsets in k-space which will manifest as a bright pixel at the center of the image. This artifact will repeat along the phase encoding direction according to the in-plane undersampling factor. To mitigate this issue, the data are first shifted by FOV/2 along the readout dimension prior to being passed through the network. In such a manner, the bright pixels will be moved to the edges of the image once the FOV/2 shift is removed following reconstruction. Performing this shifting along the readout dimension rather than the phase encoding dimension avoids any interference of the DC artifact with CAIPIRINHA-shifted SMS slices. 

\subsection{Retrospective Undersampling Experiment}
A multi-respiratory phase volumetric imaging dataset with two repetitions was acquired to test the performance of RAKI under various conditions. The first repetition was used for training, while the second was used for testing. A golden-angle 3D stack-of-stars\citep{grasp,xdgrasp} balanced steady-state free-precession (bSSFP) scan was acquired for 4.5 minutes on a 1.5T MR/linear accelerator hybrid machine (Unity, Elekta Instruments AB\citep{mrl}. Parameters include a TE/TR of 2.5/5.0 ms, flip angle of 30 degrees, imaging matrix of 256x256 and 40 partitions. The FOV was set to 340x340 mm in-plane and 240 mm through-plane. An eight-channel phased-array was used for signal reception with four channels placed anteriorly and the other four placed posteriorly. The coil sensitivity maps were estimated from the entire dataset using the Walsh method\citep{walsh}. The radial data were retrospectively sorted into 8 respiratory bins using an internally derived surrogate\citep{xdgrasp,online4d}. A slice-wise GRASP reconstruction was used to generate streak-free 4D reference images\citep{grasp}. To obtain multi-coil Cartesian data for training and testing, the GRASP images from each repetition were multiplied by the measured coil sensitivity maps and brought to k-space.

The k-space data from the respiratory phase at the end of inspiration for 4 slices each equally spaced 30 mm from their neighbors were used for a retrospective 2D acceleration experiment. To simulate CAIPIRINHA SMS imaging acquisitions, k-space data were modulated to introduce inter-slice shifts of FOV/SMS between neighboring slices. Combinations of simultaneous multislice imaging factors of up to SMS=4 and in-plane acceleration factors of up to R=4 were tested. The widely-adopted split slice-GRAPPA (SP-SG) algorithm\citep{splitSliceGrappa} with subsequent in-plane GRAPPA for each slice was used to compare RAKI with parallel imaging methods. The SP-SG and in-plane GRAPPA kernel sizes were set to 5x5 and 4x5, respectively, for all acceleration factors tested. A coil sensitivity weighted combination of the multichannel data was performed for both methods. 

The training of the RAKI networks and fitting of the SP-SG and GRAPPA kernels was performed using autocalibrating signal (ACS) k-space lines from separate slices. For reconstruction, a Monte Carlo analysis was performed. Here, Gaussian noise was added to the training and testing k-space data in 100 trials. From these data, mean structural similarity indices (SSIM) between the reconstructed and reference images were calculated\citep{ssim}. Due to the non-linearity of the RAKI reconstruction, unfortunately no meaningful g-factor can be derived from the data despite the use of Monte Carlo trials\citep{raki,csReview,parallelImagingNoise}. The SSIM values were compared between the SP-SG and RAKI methods statistically using paired t-tests for each acceleration factor. P-values beneath a significance level of 0.05 were considered statistically significant. 

\subsection{Testing Robustness to Respiratory Motion}
For RAKI to be applicable to cine imaging for respiratory motion monitoring during MR-gRT, the CNN weights learned from the reference data must be able to faithfully reconstruct images of the same slices at different respiratory phases. To test this capability, the models trained with data from the end of inspiration for the SMS=2 and R=3 case above were applied to data from the end of expiration. 

\subsection{Prospectively Accelerated SMS Phantom Experiment}
A prospectively accelerated phantom scan with SMS=2 and R=3 was performed on a Siemens 3T Verio (Siemens Healthineers, Erlangen, Germany) to test the SMS RAKI reconstruction framework under more realistic conditions. The single-band RF pulse in the fast low angle shot (FLASH) pulse sequence was replaced with a dual-band RF pulse that excites slices spaced 5 cm apart. A 12-channel head coil was used for signal reception. The data were acquired with a CAIPI FOV/2 interslice shift. A separate calibration scan with SMS=2 was performed with R=1, identical CAIPI FOV shifts, and an extended (2x) phase FOV to achieve a complete slice separation as seen in Glover (1991)\citep{pomp}. The extended FOV of each individual coil image was cropped and brought back to k-space to be used for RAKI and parallel imaging training. 

\subsection{Prospectively Accelerated Simultaneous Orthogonal Plane Imaging}
The simultaneous orthogonal plane imaging (SOPI) pulse sequence\citep{sopi} has been shown to provide utility in improving target tracking capabilities in 3D and also for dose accumulation in the presence of motion\citep{sopi4d}. Simultaneous sagittal and coronal (SMS=2) cine RF-spoiled gradient echo imaging with R=3 in-plane acceleration was tested in the abdomen of a free-breathing healthy volunteer at 3T. The slices were prescribed to intersect the liver and right kidney. Imaging parameters include TE/TR of 3.58/6.57 ms, flip angle of 10 degrees, FOV of 340x340 mm, matrix size of 128x128, and slice thickness of 10 mm. For signal reception, 21 RF receive coils placed anteriorly and posteriorly were used. Eighty frames were acquired at an acquisition frame rate of 3.62 Hz. The total acquisition time was approximately 22 seconds. Parallel imaging (SP-SG + GRAPPA) and RAKI were used to reconstruct the time series. Low-resolution calibration data (32 phase encoding lines) were obtained from each individual slice as part of a pre-scan. The parallel imaging and RAKI weights were fit from the calibration data and applied to the uniformly undersampled data with the same parameters as the retrospective acceleration experiment. 

Since these data were prospectively accelerated, no gold-standard reference images can be used to compare the quality of the PI and RAKI reconstructions quantitatively. Thus, only qualitative comparisons of the image quality between the two methods can be performed. For a fair comparison of the motion resolving capabilities of SOPI to PI methods, image intensity versus time projections were plotted from the cine data within regions outside of the severe noise enhancement seen in the parallel imaging reconstructions.

\section{Results}

\begin{figure}[H] 
\optincludegraphics[width=\textwidth]{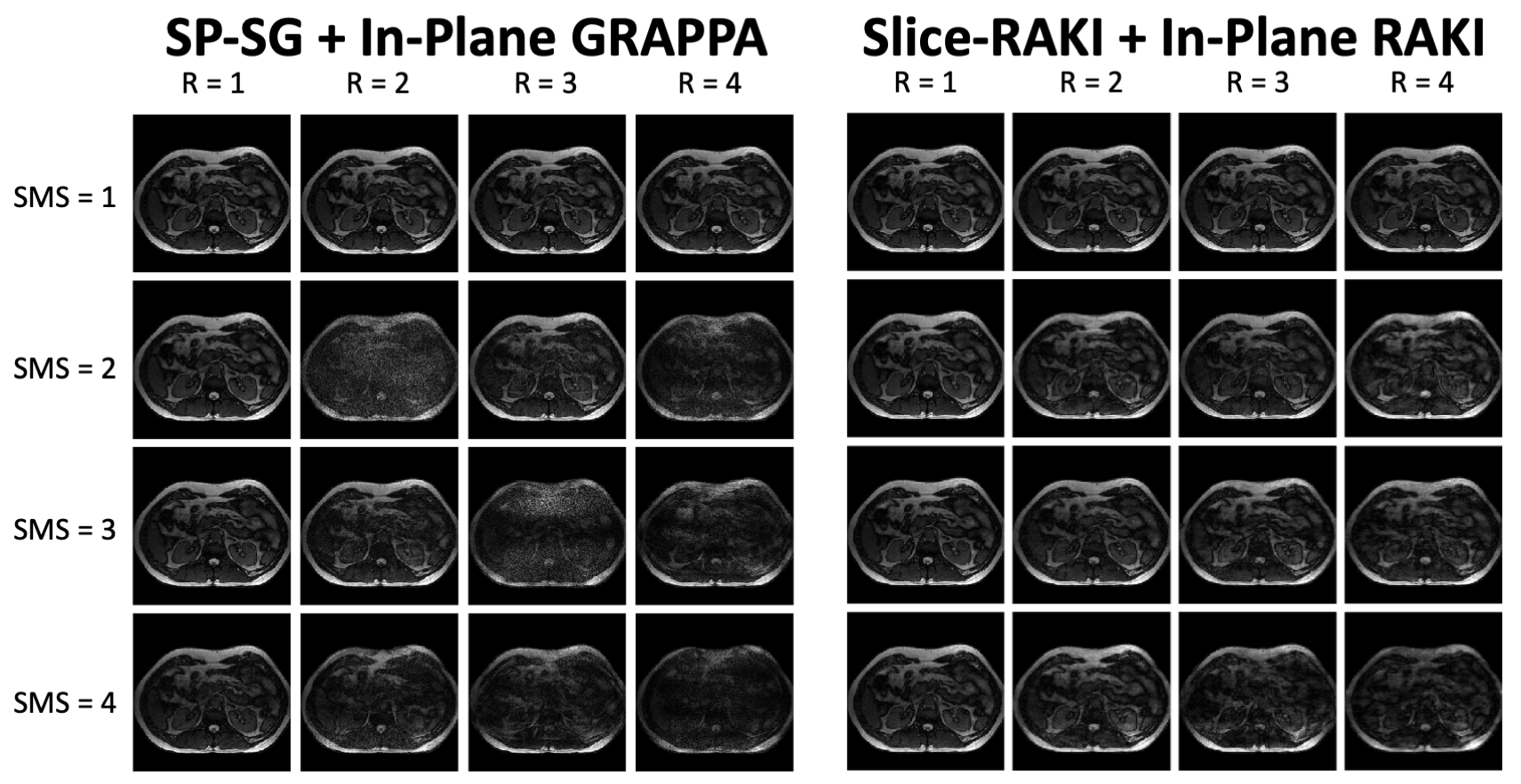}
\capt[Retrospective Undersampling Results.]{Reconstructed images at each acceleration factor for the Split slice-GRAPPA (SP-SG) + in-plane GRAPPA and slice-RAKI + in-plane RAKI methods. RAKI is able to greatly reduce the residual aliasing artifacts compared to parallel imaging. RAKI is able to perform surprisingly well for acceleration factors exceeding the number of receive coils ($N_{coils}=8$).}
\label{fig:retro}
\end{figure}

\begin{figure}[H] 
\optincludegraphics[width=12cm]{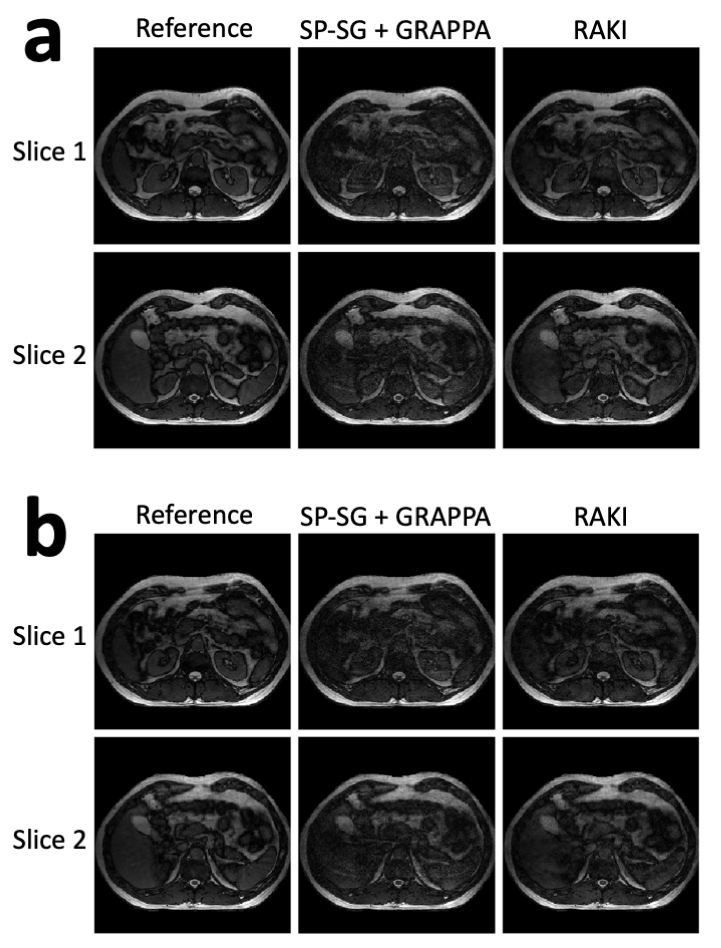}
\capt[SMS=2 with R=3 Results.]{Simultaneous multislice (SMS) factor of 2 and in-plane acceleration factor (R) of 3 reconstructions. (a) Parallel imaging weights fit and RAKI network parameters learned from end-inspiratory data and applied to end-inspiratory data. (b) Parallel imaging weights fit and RAKI network parameters learned from end-inspiratory data and applied to end-expiratory data. RAKI is able to suppress residual aliasing artifacts compared with split slice-GRAPPA (SP-SG) + in-plane GRAPPA. It is also able to perform better than SP-SG when the trained network is applied to a different motion state.}
\label{fig:sms2r3}
\end{figure}

\begin{figure}[H] 
\optincludegraphics[width=\textwidth]{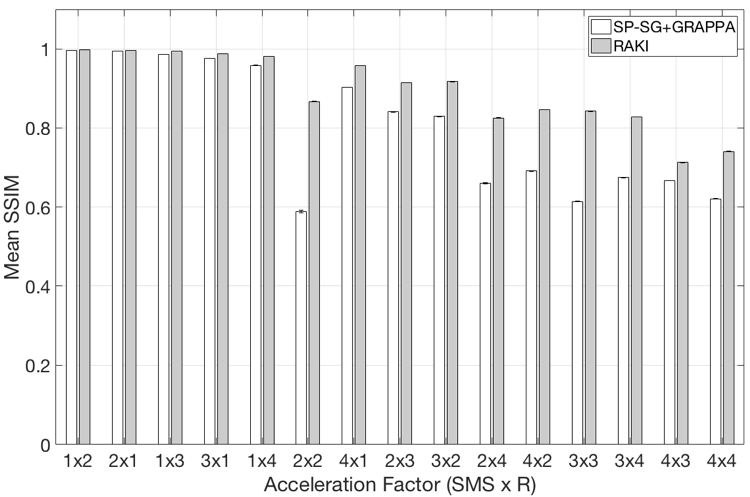}
\capt[Structural Similarity Results.]{Mean and standard deviation of the structural image similarity metric (SSIM) of split slice-GRAPPA (SP-SG) + in-plane GRAPPA and RAKI methods for varying simultaneous multislice (SMS) and in-plane (R) acceleration factors. }
\label{fig:ssimfig}
\end{figure}

Reconstructed images from the retrospective acceleration experiment are shown for both SP-SG + GRAPPA and RAKI from a single trial in Figure \ref{fig:retro}. When either the SMS or R acceleration factors are equal to one, the images from the PI and RAKI methods are very similar in quality. No residual aliasing artifacts are visible. As both SMS and R increase, the PI algorithm begins to fail as evidenced by incomplete unaliasing and noise enhancement. The RAKI algorithm is able to remove residual artifacts and suppress noise relative to PI reconstructions, even at acceleration factors greater than the number of coils. A closer view of the reconstructions for both slices for the SMS=2 x R=3 acceleration factor is shown in Figure \ref{fig:sms2r3}a. In Figure \ref{fig:sms2r3}b, reconstructions fit/trained using data from the end of inspiration and applied to data at the end of expiration are shown. This was done to test the robustness of RAKI to differences in anatomical position due to respiratory motion. RAKI is able to suppress noise and residual aliasing artifacts in the reconstructed images compared with the SP-SG+GRAPPA, even when anatomical deformations between the training and testing data are present. The mean SSIM values with standard deviations in all tested acceleration factors are plotted in Figure \ref{fig:ssimfig}. The SSIM values of images reconstructed with RAKI were significantly higher than those of images reconstructed with parallel imaging methods at all acceleration factors.

\begin{figure}[H] 
\optincludegraphics[width=\textwidth]{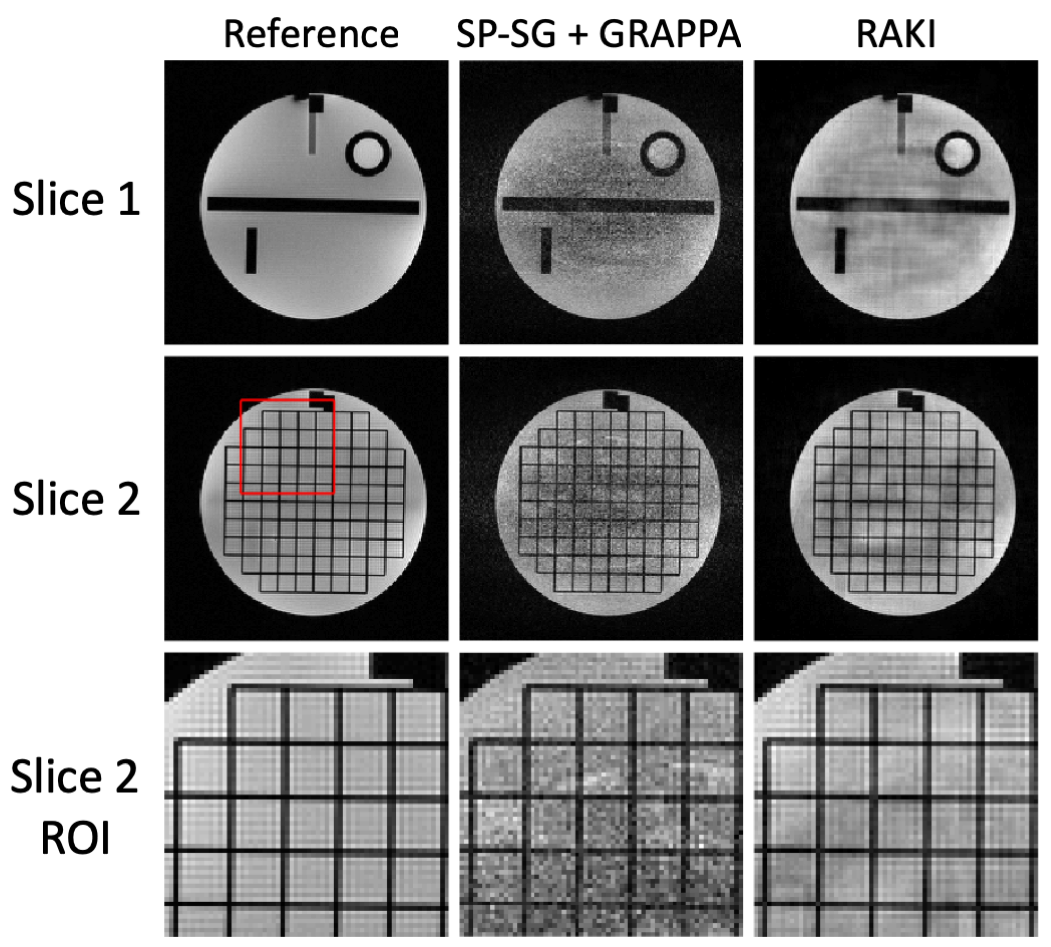}
\capt[Phantom Results.]{The parallel imaging (SP-SG + GRAPPA) and RAKI images were reconstructed form a prospectively accelerated SMS=2 and R=3 FLASH experiment. RAKI greatly reduces the amount of noise in the images compared with SP-SG with GRAPPA. The magnified region of interest (ROI) from the second slice is shown in the bottom row. This ROI shows that RAKI images maintain sharp boundaries within the grid of the structural phantom while minimizing noise enhancement.}
\label{fig:acr}
\end{figure}

\begin{figure}[H] 
\optincludegraphics[width=\textwidth]{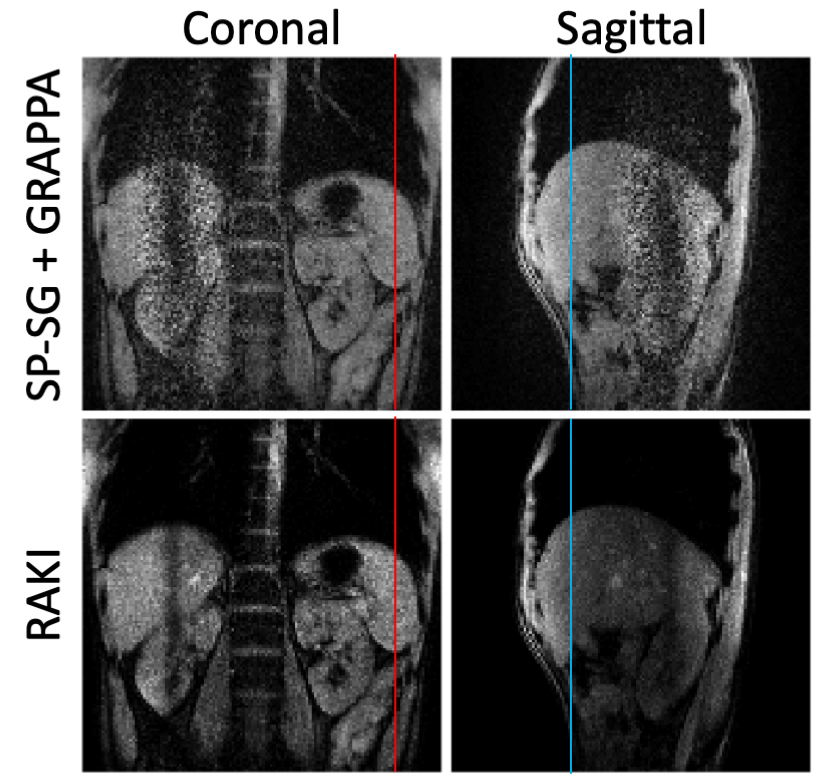}
\capt[SOPI Results.]{Reconstruction results for the simultaneous orthogonal plane imaging (SOPI) experiment. For the parallel imaging (i.e. SP-SG + GRAPPA) reconstruction, the separated orthogonal slices are contaminated by high levels of noise near the intersection of the slices. RAKI k-space interpolation yields images with less apparent noise enhancement than the parallel imaging reconstruction. Intensity projections over time were extracted from the lines shown on the images.}
\label{fig:sopi}
\end{figure}

\begin{figure}[H] 
\optincludegraphics[width=\textwidth]{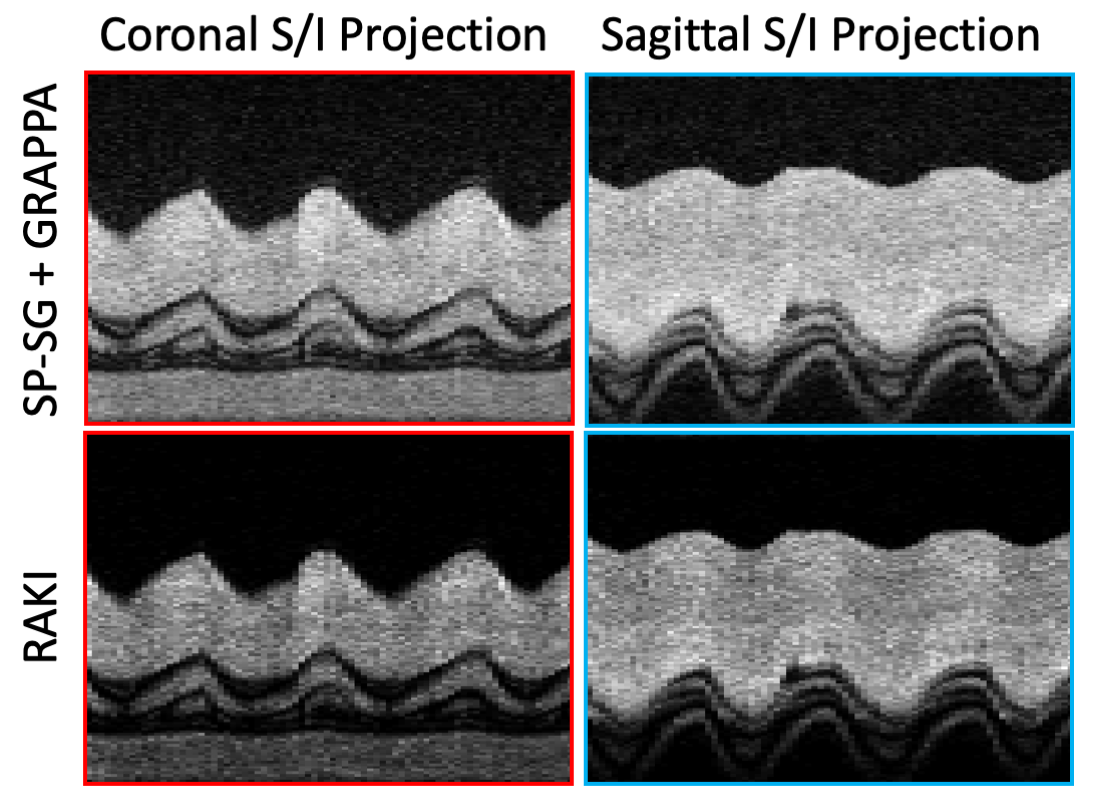}
\capt[SOPI Projections.]{Image intensity projections taken from the lines shown in Figure \ref{fig:sopi}. Each projection is shown over a time window of 22 seconds. RAKI is able to produce nearly identical projections to the SP-SG + GRAPPA reconstruction in the presence of respiratory motion.}
\label{fig:proj}
\end{figure}

Images from the prospectively accelerated SMS FLASH phantom experiment are shown in Figure \ref{fig:acr}. As expected, the parallel imaging reconstructions are highly corrupted by noise due to the high total acceleration factor ($SMS\cdot R=6$) compared to the twelve receive coils used. RAKI was able to visually improve the SNR of the reconstructed images. The RAKI images contain slowly varying intensity variations coinciding with the areas most corrupted by noise in the parallel imaging reconstructions. 
The parallel imaging and RAKI reconstructions for a frame of the SOPI acquisition are shown in Figure \ref{fig:sopi}. The SP-SG + GRAPPA images are highly contaminated by noise, particularly near the intersection of the two slices (i.e. where the coil sensitivities are most similar between the orthogonal slices). The RAKI k-space interpolation provides, visually, images of much greater quality. For example, the partial saturation band occurring due to the intersection of the two slices is not distinguishable in the parallel imaging reconstruction. Whereas in the RAKI reconstruction, this saturation band is easily identifiable. The image intensity versus time projections from the sagittal and coronal slices for parallel imaging and RAKI are shown in Figure \ref{fig:proj}. Over the course of the 22 second free-breathing scan, RAKI is able to resolve respiratory motion as well as the parallel imaging reconstructions while simultaneously providing the improved image quality demonstrated in Figure \ref{fig:sopi}.

\section{Discussion}

Cine imaging for motion monitoring during MR-guided radiation therapy demands fast frame rates, significant slice coverage, high soft tissue contrast, and image quality sufficient for structure segmentation. When acceleration factors are increased to improve temporal resolution or spatial coverage, image quality is degraded with conventional parallel imaging algorithms. In this study, RAKI was extended to perform similar k-space interpolations performed by SMS parallel imaging algorithms. RAKI outperformed conventional parallel imaging algorithms in terms of structural similarity to a reference image.  

CAIPIRINHA FOV shifts were introduced during testing of slice-RAKI to help improve image quality. When in-plane acceleration factors equal to the number of SMS slices are used, and when the inter-slice FOV shifts are FOV/SMS, the shifts are essentially removed from the acquired data. Thus, at SMSxR values of 2x2 and 3x3, the parallel imaging algorithm performed poorly as expected. RAKI was able to reconstruct surprisingly high-quality images in these cases. To help improve the quality of reconstructions at 2x2 and 3x3, different shifting patterns other than those which produce FOV/SMS shifts can be utilized. For example, FOV/3 shifts can be employed for 2x2 acceleration. With such an acquisition, an equal number of phase encoding lines would be acquired for each CAIPI modulation state. 
Currently, RAKI is implemented to interpolate in-plane and through-plane accelerated Cartesian k-space data. The extension to the non-Cartesian case is not trivial, since multiple networks would need to be trained for each “segment” of k-space due to the non-uniform sampling. If similar segmentation of k-space for radial GRAPPA would be employed for radial RAKI, 64-128 separate networks would need to be trained. This would take a significant amount of time. Furthermore, with such small segment sizes, there would not be much reference data available to perform the network training. The use of existing deep learning methods to remove streaks from coil-combined undersampled radial images may be better suited to solving the non-Cartesian imaging problem than RAKI. Variable density Cartesian acquisitions could be handled more easily than radial acquisitions. In this case, a network for each in-plane acceleration factor would be required. Typical variable density acquisitions only include two acceleration factors: an acceleration factor of R=1 near the center of k-space, and one other uniform acceleration factor at the k-space periphery. 

The choice to use a two-step RAKI implementation (i.e. slice and in-plane) was empirical. A different implementation of RAKI similar to 2D-SENSE-GRAPPA\citep{roSenseGrappa,senseGrappa2d} was tested. While the training was completed faster because the slice and in-plane unaliasing was performed in the same step in this implementation, it did not achieve as robust of results as the two-step RAKI. Further testing and optimization of the network architecture in this case may be worth investigation.

On a single 2.8 GHz CPU, the RAKI networks took up to 10 minutes to train. They could, however, be applied relatively quickly. In the SMS=2 and R=3 case with 21 RF coils, for example, each model was applied in under 0.25 seconds per frame. A significant improvement in computation speed is expected with the use of a graphics processing unit. The waiting time for training the network in the context of MR-guided radiation therapy may not be too large of a disadvantage. Currently, it can take on the order of 5-10 minutes to adapt a treatment plan based on pre-treatment volumetric scans acquired of the patient. Therefore, the brief (< 2 seconds) calibration scan required for RAKI can be performed first, and the training can be immediately started. With such a workflow, the trained networks will be ready for real-time application as soon as the adaptive treatment plan has finished calculating. 

The comparison of RAKI was limited to non-regularized parallel imaging algorithms. This choice was made because of the target application for RAKI. Computationally demanding compressed sensing reconstructions, which iteratively solve for the reconstructed image while imposing some form of sparsity constraint, are too time consuming for real-time motion monitoring MR-gRT applications. The reconstructions must be rapid, such as a forward pass of the undersampled k-space data through a pre-trained RAKI network. If the use of RAKI for accelerating diagnostic MR scans is desired, then a comparison to more robust algorithms such as ESPIRiT\citep{espirit}, SAKE\citep{sake}, or P-LORAKS\citep{ploraks} should be performed. 

Phase constrained parallel imaging algorithms make use of conjugate symmetry in k-space to further encode the data and reduce noise enhancement\citep{phaseConstrainedMR}. Furthermore, in SMS acquisitions, phase differences between simultaneously excited slices can be exploited to improve image quality. Potentially, a similar approach to the virtual conjugate coil method\citep{vcc,smsPhaseConstrainedMR} could be used in conjunction with RAKI. The virtual coil data could be passed into the network while the interpolated real channels could be output from the network. This will make for an interesting avenue of further research. 

A unique scan-specific deep learning approach to undersampled MR image reconstruction for cine imaging on the MR-Linac was presented. It shows promise to be capable of improving the amount of spatial coverage obtained while maintaining temporal resolution compared with conventional parallel imaging techniques. The use of RAKI in conjunction with the 4D-SOPI method\citep{sopi4d} with simultaneous 2D cine and 4D imaging for dose accumulation on the MR-Linac will be investigated.



\bibliography{references}




\section{Figures and Tables}

\end{document}